\documentclass[aps,pre,twocolumn,superscriptaddress,floatfix]{revtex4-2} 

\usepackage{graphicx}   
\usepackage{amsmath}    
\usepackage{amssymb}    
\usepackage{hyperref}   
\usepackage{bm}         
\usepackage{subfigure}
\begin{document}

\title{From Disorder to Design: Entropy-Driven Self-Organization in an Agent Based Swarming Model and Pattern Formation}

\author{Vinesh Vijayan}
\author{Karpagavalli K}
\author{Sandhiya Jenifer J}
\author{Prakash R}
\affiliation{Department of Science \& Humanities, Rathinam Technical Campus, Coimbatore, India}
\email{vinesh.phy@gmail.com}  
\date{\today}

\begin{abstract}
This letter seeks to illuminate the profound connection between complexity, self-organization, emergent behaviour, pattern formation, and entropy—concepts that are foundational to understanding our universe. By examining these ideas through the lenses of physics, information theory, and nonlinear dynamics, we uncover a fascinating narrative. Starting with a random cluster of particles possessing distinct internal properties, we activate their interactions and observe the emergence of intricate patterns over time. This journey reveals a transition from unlikely to more probable states. At extreme parameter values, the system showcases stunning patterns and turbulent motions—remarkable emergent behaviour propelled by entropy and the dynamic exchange of mutual information. Engaging with probability theory helps us to unveil this intricate connectivity, demonstrating not only its significance but also its potential to reshape our understanding of complex systems.

\end{abstract}

\maketitle
\section{Introduction}
Complex systems, found everywhere, are characterized by the emergence of global behaviours. Individual agents, following simple rules, interact with each other locally, exchanging information with their nearest neighbours and the surroundings\cite{And}\cite{Sim}. These interactions give rise to global behaviours that are not exhibited by the individual agents, all without a central organizer. Spatial patterns, synchrony, or coordinated functional capabilities are some of the global properties observed in these systems\cite{Swe}\cite{Srd}\cite{Bri}. A swarming system, often modelled as a collection of mobile oscillator, is one good example for studying emergence and self-organization that finds application in many natural and artificial systems. It opens up an understanding of swarm intelligence, free will, and consciousness\cite{Dav}\cite{Sie}\cite{Ess}. One can understand how efficiently nature organizes and distributes energy as part of a free energy minimization process in all the above mentioned global behaviours\cite{Kau}\cite{Axe}\cite{Wol}.\\
\\
In our efforts to understand complex systems and self-organization, entropy emerges as a crucial measure of energy effectiveness\cite{Bol}\cite{Sha}\cite{Jay}. From an energy perspective, entropy gauges the efficiency or usefulness of a given amount of energy, suggesting that energy is most effective when concentrated and less so when dispersed. As per the second law of thermodynamics, the total entropy of the universe is increasing, leading towards a state of greater disorder\cite{Lan}\cite{Cla}.
\begin{equation}
    dS \geq \frac{dQ}{T}
    \label{E1}
\end{equation}
where, $dS$ is the change in entropy expressed in (J/K), $dQ$ is the amount of heat energy expresed in (J) and $T$ is the absolute temperature at which heat transfer occurs expressed in(K).  However, the universe is considered a closed system. In contrast, complex systems are open systems, capable of exchanging energy with their environment. This unique feature allows the system to self-organize into structured patterns without a central organizer.  One can view entropy as a measure of disorder or randomness. As stated before, all the processes in the universe occur in the increasing direction of disorder or entropy. At zero kelvin for a pure crystalline substance, the entropy is zero since there is no uncertainty in the states of the particles.
\begin{figure}[h]
    \centering
    \includegraphics[width=0.8\linewidth]{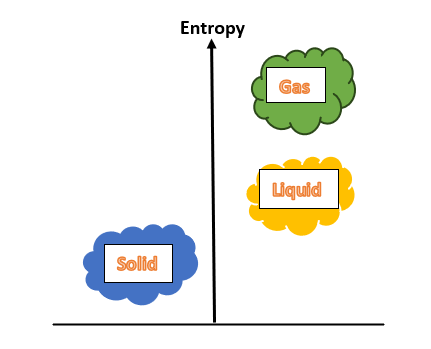}  
    \caption{Entropy scale of matter.}
    \label{F1}
\end{figure}
Understanding energy and entropy is pivotal in deciphering complex systems. Energy minimization, a sign of stability, is when the system tends towards a low-energy state. The optimization of free energy leads to intriguing phenomena such as self-assembly, phase transition, and spontaneous processes. Unlike traditional states of matter, soft matter like colloids, liquid crystals, and biological membranes exhibit reversible organization, which is dynamic and flexible. The interplay of environmental impacts, entropy and energy minimization can show how complex structures emerge from local interactions. Swarming systems, composed of mobile oscillators, are complex and can effectively model the dynamics of both natural and artificial systems\cite{Kol}\cite{Sar}\cite{Ana}\cite{Oke1}\cite{Oke2}\cite{Hon}\cite{Oke3}. The simulation results discussed in various studies are particularly fascinating, offering profound insights into the underlying patterns and mechanisms that shape nature's grand design\cite{Lee}\cite{Oke4}\cite{Yoo}\cite{Sar2}. This paper aims to establish this relation by modelling and simulating an agent-based swarming system, a potential game-changer in understanding soft matter dynamics.\\
\\
This work is the continuation of our own research as mentioned reference\cite{Vij3}. The swarming system is modelled by sewing Kuramato-type phase oscillations\cite{Kur} with spatial dynamics represented by cyclically symmetric Thomas oscillators\cite{Tho}\cite{Vij1}\cite{Vij2}. This innovative approach to agent-based swarming modelling will then be simulated for the different parameter regimes to capture the important dynamic features. At each instant of time, including the study states, the locations and phases of the particles will be specified. The entropy and randomness of this data set will be analysed with the help of probability theory to establish the connection between order/ disorder and entropy.\\
\\
This paper is structured as follows: Section II introduces the mathematical modelling framework and outlines the probabilistic measures used in the analysis. Section III presents numerical results, examining three distinct regimes of the system parameters for two different interaction parameter choices. Section IV connects these findings to real-world observations, emphasizing their technological implications. Finally, Section V provides a comprehensive conclusion, summarizing key insights and potential avenues for research.
\section{Mathematical Modelling and Numerical Simulations}
\begin{figure}[hbt!]
      \centering
    \includegraphics[width=0.9\linewidth]{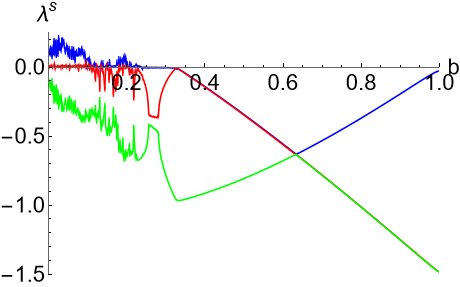}  
    \caption{Lyapunov spectrum for Thomas oscillator.}
    \label{F2}
\end{figure}

We propose a 3D swarming model of the following form with attractive and repulsive coupling\cite{Oke1}\cite{Hao}, In this model, the phase stands for some internal property of the particle, like spin or magnetic moment, an internal degree of freedom. $J$ stands for the spatial interaction strengths of particles based on their phase, and $K$ is the phase coupling constant. These two control variables, along with the system parameter, which plays a crucial role, regulate the swarm behaviour in the proposed model. By tuning the values of $J$ and $K$, one can understand different swarming states in conjunction with the damping parameter. 
\begin{equation}
\begin{aligned}
\frac{d{\bf r_i}}{dt}&={\bf f(r_i)} +\frac{1}{N}\sum_{j\neq i}^N[\frac{{\bf r_j-r_i}}{\mathit{r}}(1+JCOS(\Theta_j-\Theta_i))\\
& -\frac{{\bf r_j-r_i}}{\mathit{r}^3}]\\
\frac{d \Theta_i}{dt}&= \frac{K}{N}\sum_{j\neq i}^N \frac{SIN(\Theta_j-\Theta_i)}{\mathit{r}^2}
\end{aligned}
\label{E2}
\end{equation}
where  ${\bf f}$ is given by the cyclically symmetric Thomas system
\begin{equation}
\begin{split}
\frac{dx}{dt} = -bx +siny\\
\frac{dy}{dt} = -by +sinz \\ 
\frac{dz}{dt} = -bz +sinx
\end{split}
\label{E3}
\end{equation}
${\bf r} = [x, y, z]^T$, ${\bf v} = [\dot{x}, \dot{y}, \dot{z}]^T$, and ${\bf f} = [f_1, f_2, f_3]^T$, with all vectors belonging to $\mathcal{R}^3$. Two key characteristics of this system deserve closer examination: its symmetry under the cyclic interchange of the $x$, $y$, and $z$ coordinates, and the single parameter $b$, a crucial factor that represents the frictional damping coefficient and adjusts the system between chaotic and regular motion. The symmetry forms a feedback loop that regulates the particle's local dynamics. Parameter b, frictional damping, is a nuanced concept. A very low value implies a sparse lattice, where the particle's interaction with its environment is rare. On the other hand, a high value suggests a densely packed environment. Thomas's system, therefore, is instrumental in providing crucial information about both the particle and the environment, enriching our study. The presence of feedback loops enhances the appeal of the Thomas system for studying biological phenomena, as these mechanisms regulate the system's state variables. Therefore, it is essential to highlight how the Thomas system can model swarming behaviour, microbial motility, and self-propulsion. For instance, dynamically interacting organisms and their surroundings contribute to the production, cycling, and regulation of energy and matter, all of which involve feedback circuits. Within the Thomas system, these circuits can be either positive or negative, depending on the state variables. Ultimately, the system's three-dimensional dynamics are influenced by both the control parameter $b$  and the nature of the feedback mechanisms. By adjusting this parameter, we can study the swarming process, ranging from highly turbulent environments to highly ordered fluid flows.\\
\\
At this stage, we bring into focus the dynamics of an uncoupled Thomas oscillator. The damping parameter $b$ is the driving force behind these dynamics, a clear indicator of the system's dissipative nature. By adjusting $b$, we can guide the dynamics from fixed points at higher values to chaotic behaviour at lower values ($b = 0$). The visual aids of FIG.{\ref{F2}}, the  Lyapunov spectrum, provide a tangible representation of these complex dynamics. For $b > 1$, the system is drawn towards a single attractive fixed point: the origin. However, between $b = 1$ and $b = 0.328$, the system presents two attractive fixed points, adding a layer of complexity to the dynamics. At the critical value of $b = 0.328$, the system experiences a Hopf bifurcation, leading to the emergence of a limit cycle. As $ b$ further decreases to $ 0.208$, the system embarks on a journey towards chaotic behaviour, with quasi-periodic windows punctuating the chaotic regime\cite{Tho}.\\
\\
\\
In this work, we conduct numerical experiments on swarmalators modelled by Equation (\ref{E2}). We use the RK4 method to solve the coupled differential equations. All swarmalators start in a cubic box with side length two, and their phases are initially drawn uniformly at random from $[\pi, -\pi]$. We then observe and analyse the subsequent collective dynamics.  
One can differentiate the stationary states from the non-stationary states by measuring the mean velocity as follows for the 3D case:

\begin{equation}
\begin{aligned}
\mathcal{V} = \left\langle \frac{1}{N} \sum_{i=1}^N \sqrt{\dot{x}_i^2 +\dot{y}_i^2+\dot{z}_i^2+ \dot{\Theta}_i^2} \right\rangle_t
\end{aligned}
\end{equation}
Here, the time average is taken. A finite non-zero value of the mean velocity indicates that the swarmalators are moving in space, and their phases vary in the interval $[0,2\pi)$. For a stationary state, $\mathcal{V}=0$.\\
\\
The mobility of the swarm in the steady state is analysed by examining the motion of the center of mass of the particles. A stationary swarm is characterized by a fixed centre of mass. Any variation in the potion  of the center of mass indicates movement, resulting in a non-zero rms displacement of the centre of mass.
Assuming same mass, for N particles, the center of mass $R_{CM}$ is given by the average position of all the particles.
\begin{equation}
\mathbf{R}_{\text{CM}} = \frac{1}{N} \sum_{i=1}^{N} \mathbf{r}_i
\end{equation}
where ${\bf r_i}=(x_i,y_i,z_i)$ is the position of the $i^{th}$ particle. The component form of the above equation is given by
\begin{equation}
X_{\text{CM}} = \frac{1}{N} \sum_{i=1}^{N} x_i, \quad
Y_{\text{CM}} = \frac{1}{N} \sum_{i=1}^{N} y_i, \quad
Z_{\text{CM}} = \frac{1}{N} \sum_{i=1}^{N} z_i
\end{equation}
The rms displacement of the centre of mass is giveny
\begin{equation}
\mathcal{D} = \sqrt{ \left\langle \left| \mathbf{R}_{\text{CM}}(t_f) - \mathbf{R}_{\text{CM}}(t_0) \right|^2 \right\rangle }
\end{equation}
If from the transient time $t_0$ to the steady state $t_f$, the time-averaged rms displacement is non-zero, which marks the motion of the centre of mass.

\subsection{Entropic Measures of the Swarming States}
When the probability of a specific event is either zero or one, the system is devoid of uncertainty. This results in an entropy of zero, indicating the absence of new information. However, when the outcome is uncertain, the entropy reaches its maximum. Shannon's entropy, a key concept in information theory, quantifies the amount of missing information needed to completely define the state of a system. Its mathematical representation is
\begin{equation}
H = - \sum_{i} p_i \log p_i
\label{E4}
\end{equation}
and measures the information content in a given data. In a swarming system with many interacting particles, one can specify the positions and momenta of these particles at each instant. The Hamiltonian of the system specifies its energy. Shannon entropy can quantify the randomness and uncertainty in this large data set. Boltzmann entropy, a thermodynamic concept, is a measure of the disorder of a system given by
\begin{equation}
S = k_B \ln \Omega
\label{E5}
\end{equation}
where $\Omega$ is the number of accessible micro-state of the system, $k_B$ the Boltzmann constant.  For an equally probable system the probability $p_i = \frac{1}{\Omega}$ the Shannon entropy is given by
\begin{equation}
H = - \sum_{i} \frac{1}{\Omega} \log \frac{1}{\Omega} = \log \Omega
\label{E6}
\end{equation}
As modelled above, the study of how information is stored and transmitted in an active matter system is not just interesting, but also significant. It is a key area of research that has the potential to impact various fields.
\\
Another probabilistic measure is the joint probability between two or more variables simultaneously. For two variables, it is defined as
\begin{equation}
H(X, Y) = - \sum_{x \in X} \sum_{y \in Y} p(x, y) \log p(x, y)
\label{E7}
\end{equation}
where $p(x, y)$ is the joint probability distribution of variables X and Y. It measures the randomness in two variables together. Mutual information captures how efficiently information transformation between interacting multi particle systems happens. For two variables, it is defined as
\begin{equation}
I(X; Y) = H(X) + H(Y) - H(X, Y)
\label{E8}
\end{equation}
and quantify the amount of information shared between two variables, reducing the mystery over the other. The maximum mutual information occurs when two variables entirely depend on each other.
\begin{equation}
I_{\text{max}} = \min(H(X), H(Y))
\label{E9}
\end{equation}
Probability, entropy, joint entropy and mutual information, concepts from information theory, plays crucial role in understanding swarming behaviour since it involves interaction among many interdependent agents.
\section{Swarming Results}
Out of the many combinations of the parameters of the interaction $J$ and $K$, we specifically focus on($J> 0, K=1$) and ($J >0, K=-1$), which stand for synchronization and desynchronization regime of the phase in the swarming dynamics. We have studied these under three specific values of the damping parameter $(b-1.2, 0.5, 0)$. Our attention has been on the dynamics around the single stable fixed point at the origin for $b=1.2$, the two stable fixed points at $b=0.5$, and the conservative dynamic regime at $b=0$, each exhibiting rich and profound dynamical behaviours.
We discovered distinct states that had not been previously reported while modelling the complex system as a swarming model and is consolidated in Table(\ref{T2}).
\begin{table*}
\centering
\caption{New swarming states.}
\begin{tabular}{ |p{5cm}|p{6cm}|p{3cm}|  }
\hline
\multicolumn{3}{|c|}{  Dynamic Swarming states} \\
\hline
Asymptotic States & Parameter Values & Abbrevations \\
\hline
1.Dynamic Sync &$b>>0, J>0, K=1,\mathcal{V}\neq0, \mathcal{D}(t) =0$& DS \\
2.Dynamic Async &$b>>0, J>0 ,K=-1,\mathcal{V}\neq0, \mathcal{D}(t) =0$  & DAS\\
3.Turbulent Sync &$b \simeq 0,J>1,K=+ve,\mathcal{V}\neq0, \mathcal{D}(t)\neq0$& TS\\
4.Turbulent Async &$b \simeq 0, J=1,K=0,-ve,\mathcal{V}\neq0, \mathcal{D}(t) \neq0$&TAS\\ 
\hline
\end{tabular}
\label{T2}
\end{table*}

\subsection{Dissipative regime - Swarming states for $b>1$}
\begin{figure*}[hbt!]
    \centering
    \subfigure[DS(t=1250)]{\includegraphics[width=0.3\linewidth]{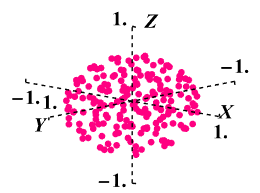}}
\hfil
     \subfigure[DAS(t=1320)]{\includegraphics[width=0.3\linewidth]{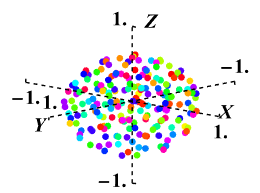}}\vfil
    \subfigure[DS(t=1250)]{\includegraphics[width=0.3\linewidth]{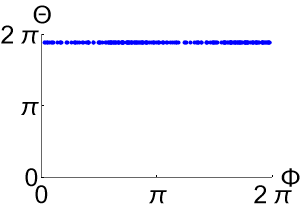}} 
\hfil
       \subfigure[DAS(t=1320)]{\includegraphics[width=0.3\linewidth]{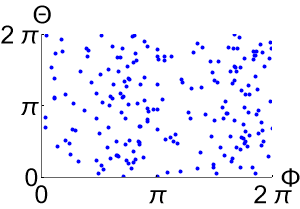}}  
\caption{ {\bf Top panel}: Swarming states in 3D Space. {\bf Bottom panel}: The $(\Phi, \Theta)$ distribution space for different swarming states. Simulations are done for $N=200$ for $T=1500 $ time units with $dT=0.01$ and $b=1.2$. (a,c)DS $(J,K)=(0.1,1)$,   (b,d) DAS  $(J,K)=(0.1,-1)$. } 
 \label{F3}
\end{figure*}

\begin{figure*}[hbt!]
    \centering
    \subfigure[DS]{\includegraphics[width=0.48\linewidth]{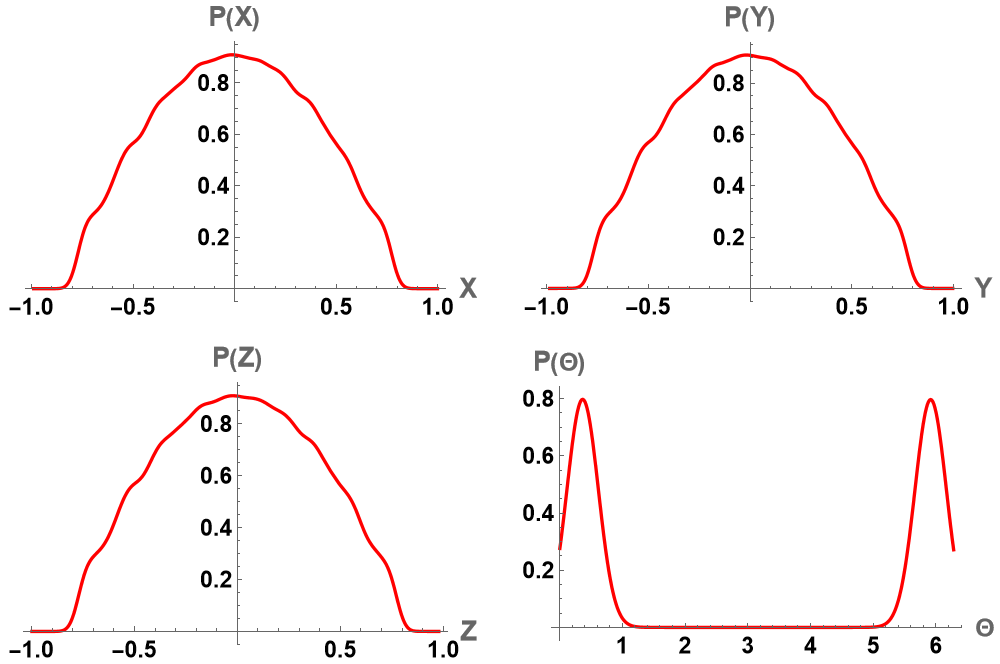}}
\hfil
     \subfigure[DAS]{\includegraphics[width=0.48\linewidth]{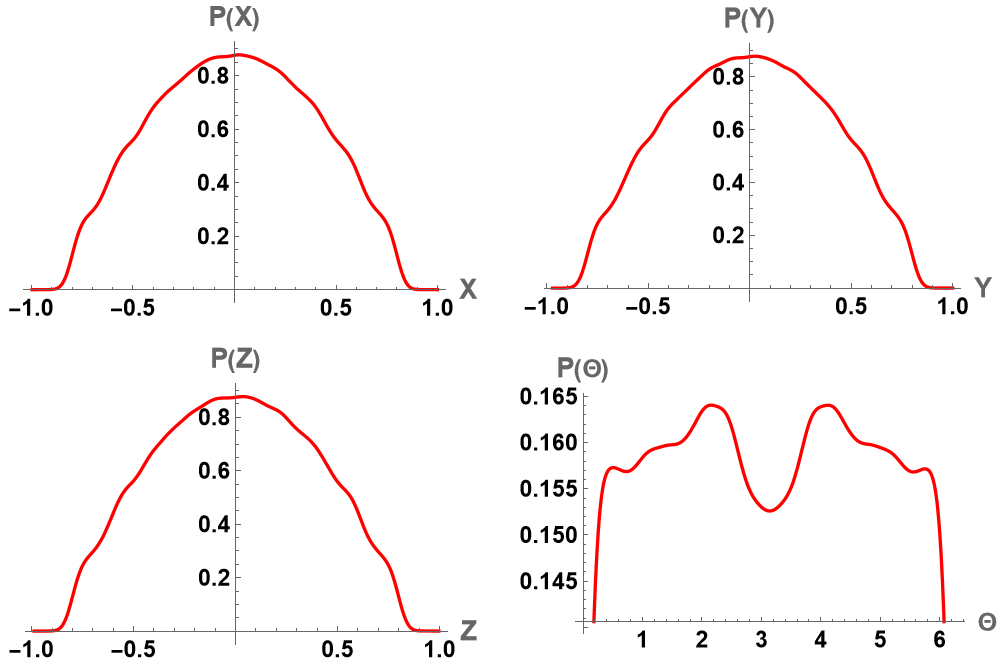}}  
\caption{  The probability distributions of space and phase variables  (a) DS (b) DAS  case. } 
 \label{31}
\end{figure*}

\begin{figure*}[hbt!]
    \centering
    \subfigure[DS]{\includegraphics[width=0.48\linewidth]{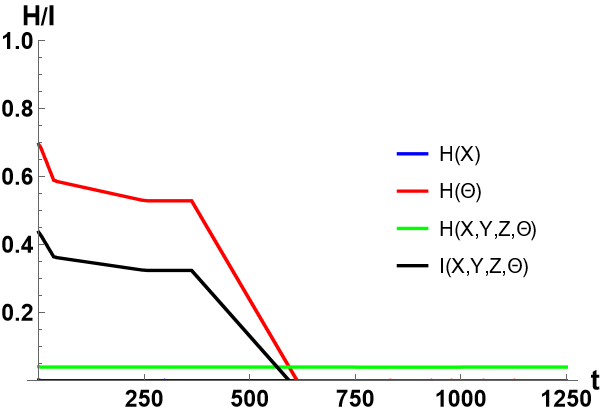}}
\hfil
     \subfigure[DAS]{\includegraphics[width=0.48\linewidth]{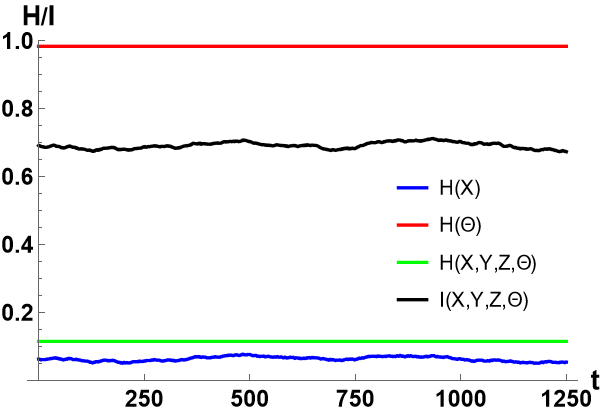}}  
\caption{ Shannon entropy for space(Blue) and phase(Red) variables, joint entropy(Green), Mutual information(Black) plots for (a) DS (b) DAS  } 
 \label{F311}
\end{figure*}

For $b>1$, indicating a high control parameter, the uncoupled Thomas oscillator has a single equilibrium point—an attractive fixed point at ($x^*=y^*=z^*=0$). In this state, the swarmalators exhibit collective behaviour that evolves into a circularly symmetric distribution in three-dimensional space centred at the fixed point with a specific orientation. The viewpoint is adjusted to emphasize this circular symmetry clearly.  At elevated damping levels, the system exhibits several distinct behaviours. Notably, static swarming states do not emerge under these conditions. For example, when $J>0$ and $K=1$  , the system undergoes dynamic synchronization (DS) [FIG.\ref{F3}(a)], in contrast to the static swarming (SS) state previously reported in \cite{Oke1}. In the SS regime, swarmalators remain stationary in space while their phases asymptotically synchronize to a uniform value. However, in the DS state observed here, particles engage in persistent quivering motion within the circular disk with a small variance, with their phases oscillating between two fixed values. This dynamic interaction gives rise to an intricate pattern within the disk, resembling an irregular hexagonal structure manifestation of the underlying complex  interactions. The corresponding phase and spatial relationship is given in FIG.\ref{F3}(c). Particles at different spatial orientation shares the same phase showing lack of space dispersion.\\
\\
As the system evolves beyond its initial transients, the swarmalators attain complete phase synchronization. This alignment indicates that the particles adapt certain internal properties to sustain coherence. The observed dynamics likely arise from the persistent local competition between phase-dependent spatial interactions and damping, which drives a continuous cycle of energy dissipation and replenishment. The non-zero value of  $\mathcal{V}$  further validates the system's dynamic nature. However, despite the internal motion of the particles, the circular disk as a whole remains stationary.\\
\\
The Dynamic async state (DAS) is obtained for $(J, K)=(0.1,-1)$ and represents complete de-synchronization of the phase. There is maximum phase-space dispersion. Even then, we get a circular pattern where the particles arrange from the centre to the outward direction isotropically. This situation is depicted in FIG.\ref{F3}(b,d).
From a mathematical standpoint, the formation of a circular boundary signifies that conditions are isotropic, ensuring uniformity in all directions. Within this framework, the swarm expands symmetrically outward from a central point—the stable fixed point at $(0,0,0)$. This outward movement exemplifies the swarm’s inherent capacity for self-organization, wherein large-scale interactions give rise to stable and structured patterns. Much like the formation of bubbles in three-dimensional space, which minimize surface tension by adopting a spherical shape, the circular symmetry observed here adheres to the principle of maximizing area while minimizing perimeter, thereby optimizing energy distribution. This symmetry also mitigates edge forces, facilitating efficient expansion while maintaining energy regulation.\\
\\
The model presented not only elucidates the mechanics of swarm movement but also reveals deeper connections to fundamental principles of energy conservation and symmetry, which are widely applicable across biological and physical systems. While the overall structure of the swarm remains circular due to the isotropic spreading, localized interactions give rise to transient geometric patterns, notably hexagonal arrangements. Hexagonal tessellations are prevalent in nature, as they provide an optimal packing strategy that maximizes spatial efficiency while minimizing excess gaps. However, the transient nature of these patterns stems from the dynamic interplay between localized crowding effects and the uniform outward expansion of the swarm. This tension between order and instability induces a characteristic quivering motion within the circular boundary, encapsulating the intricate balance between organization and emergent fluctuations.\\
\\
By plotting the probability density function, we reveal the crucial behaviour of the system. All the position space state variables follow a bell-shaped distribution that is symmetric and centred around the origin. The particles prefer to stay around the origin without dispersing and follow a thermal distribution. However, the phase probability density function shows a distinct pattern with two peaks, one near zero and the other near six, with no other phase values allowed. This unique behaviour, with its two distinct peaks, is of utmost importance as it indicates a preferred angular orientation, or in other words, bi-stability in the phase coordinates, as the two clusters indicate. The coexistence of two phases within the same environmental conditions and interactions suggests that the swarm can settle into any one of the two depending on the initial conditions. One can connect this with the circulating and translatory motion within the circular disc. For the DAS case, except for the phase probability density function everything remains the same. But the phase probability distribution function shows a spread in $\Theta $ values showing the allowed values of $\Theta$ (almost a uniform distribution) and hence hi-lights the phase disorder as per the interaction constants. These scenarios are demonstrated on the left and right panels of FIG.\ref{31}\\
\\
The Shannon entropy and mutual information calculation for the DS state show that([FIG.\ref{F311}(a)]), initially, the system is enveloped in high uncertainty and dependencies among state variables. However, as time progresses, Shannon entropy and mutual information decrease, indicating the system's evolution towards a stable, steady state of predictability. This predictability is a reassuring sign of the system's behaviour, instilling confidence in its stability. Around $t\sim750$, the last traces of randomness are removed, and the system moves towards order. In the steady state, $H(\Theta)$ and $I(x,y,z,\Theta)$ remain zero, indicating no randomness or shared information among variables. Yet, the joint entropy function suggests the system's minimal randomness, which stabilizes above zero values and highlights constraints among variables. There is also a hint of microscopic randomness in $X$ values, with minimal variance. For the DAS case, the system is found to be highly constrained with predictable $X$ and maximum randomness in $\Theta $([FIG.\ref{F311}(b)]). There are significant dependencies between all the variables, and mutual information remains relatively high, implying that the variables share much information. The Joint entropy function is not just low, but remarkably so, indicating the strength of the constraints among variables. $ H (\Theta) $ remains one, indicating that there is maximum randomness in $\Theta $, and all possible values between $ 0-2\pi$ are possible, which results in a uniform distribution.
\\
\subsection{Dissipative regime - Swarming states for $b \in (0.328,1)$}
\begin{figure*}[hbt!]
    \centering
    \subfigure[DS(1200)]{\includegraphics[width=0.3\linewidth]{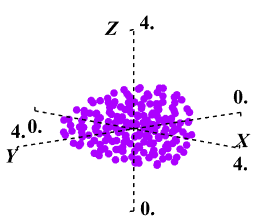}} 
    \hfil    
     \subfigure[WCAPW(1200)]{\includegraphics[width=0.33\linewidth]{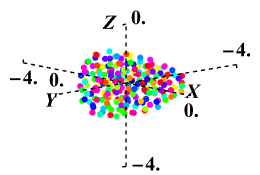}}\vfil
    \subfigure[DS(1200)]{\includegraphics[width=0.3\linewidth]{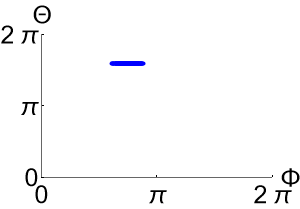}} 
    \hfil    
     \subfigure[WCAPW(1200)]{\includegraphics[width=0.33\linewidth]{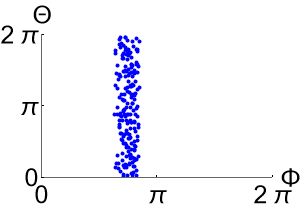}}  
\caption{{\bf Top panel}: Swarming states in 3D Space. {\bf Bottom panel}: The $(\Phi, \Theta)$ distribution space for different swarming states. Simulations are done for $N=200$ for $T=1500 $ time units with $dT=0.01$ and $b=0.5$. (a,c) DS $(J,K)=(0.1,1)$,  (b,d) WCAPW $(J,K)=(1,-1)$.} 
 \label{F312}
\end{figure*}

\begin{figure*}[hbt!]
    \centering
    \subfigure[DS]{\includegraphics[width=0.48\linewidth]{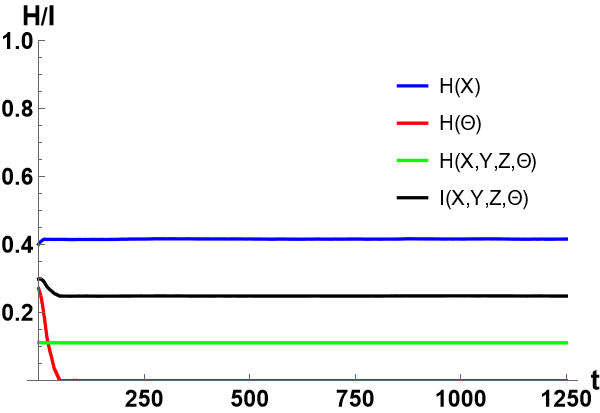}}
\hfil
     \subfigure[WCAPW]{\includegraphics[width=0.48\linewidth]{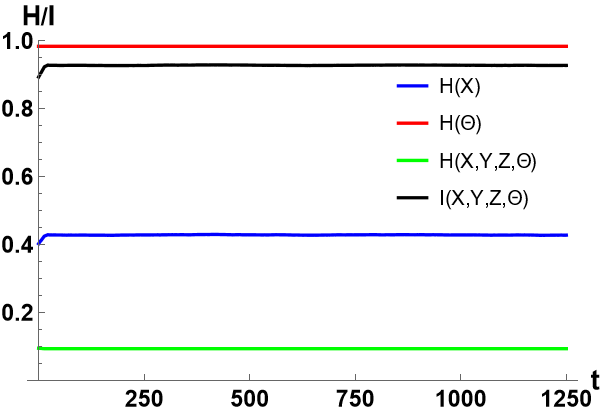}}  
\caption{  Shannon entropy for space(Blue) and phase(Red) variables, joint entropy(Green), Mutual information(Black) plots for (a) DS (b) WCAPW } 
 \label{F33}
\end{figure*}

At $b=1$, the uncoupled Thomas system undergoes a pitchfork bifurcation, resulting in the emergence of two symmetrically positioned attractive fixed points. These equilibrium points,  given by  $(x^*=y^*=z^*=\pm \sqrt{6(1-b)})$  introduce compelling collective dynamics. For the specific case of $b=0.5$, the swarmalators self-organize into clusters centred at the fixed points  $\pm(1.73,1.73,1.73)$,  eventually forming distributions with boundaries resembling a Zindler type curve—a disk and its mirrored counterpart, reflected about an axis parallel to the x-axis in three-dimensional space.  This symmetry suggests a uniform density distribution, akin to objects maintaining stability while floating in a fluid, independent of their orientation. The viewpoint is deliberately chosen to emphasize this structural symmetry. One notable difference from the previous case is the absence of the DAS state for$(J>0, K=-1)$. Instead, we have observed a weakly correlated active phase wave(WCAPW). This observation, from zero to $-1$, presents a scenario with the correlation strength gradually decreasing from a high value to a low value. The most pressing issue is the uncertainty in choosing the fixed point out of the two, which calls for further investigation. The selection of either configuration may be energetically favourable; however, the precise mechanisms driving this preference remain uncertain. It appears to be influenced by factors such as interaction strengths, the number of oscillators, and initial conditions. Numerical simulations conducted with two hundred swarmalators revealed two distinct scenarios. In cases where the swarmalators converge toward the positive fixed point, they evolve into a disk-like distribution exhibiting Zindler-type symmetry. Conversely, when the negative fixed point is favoured, the same Zindler-type symmetry governs the distribution, as illustrated in top panel of FIG.\ref{F312}. The phase space orientation curve is also interesting because we have phase synchrony for the DS, notably around specific spatial orientations. This means that the particles tend to propel. On the other hand, for the WCAPW case, there is no phase synchrony, but the spatial orientation still shows the same tendency as before. This scenario is depicted in the bottom panel of FIG.\ref{F312}.
\\
\\ 
The Zindler curve shape, a product of the swarm's ingenious optimization process, is a testament to the complexity of nature. The swarm, in its quest to maximize propulsion efficiency while minimizing disruption caused by physical constraints, engages in a complex interplay between swarm motion, surface interactions, and self-organizing behaviour. The convex boundary, a delicate balance of propulsion, friction, and particle interactions, is a marvel of natural engineering. This shape emerges from the swarm's movement away from or toward a fixed point, a process that is both elegant and efficient. The central void in the structure, a result of crowding-induced repulsion, is a fascinating example of how the swarm leaves behind a space while maintaining collective, organized movement. Within the Zindler-shaped disc, the swarm twists while keeping the underlying hexagonal tessellation, a feat that showcases the swarm's adaptability and resilience.\\
\\
\begin{figure*}[hbt!]
    \centering
    \subfigure[DS]{\includegraphics[width=0.48\linewidth]{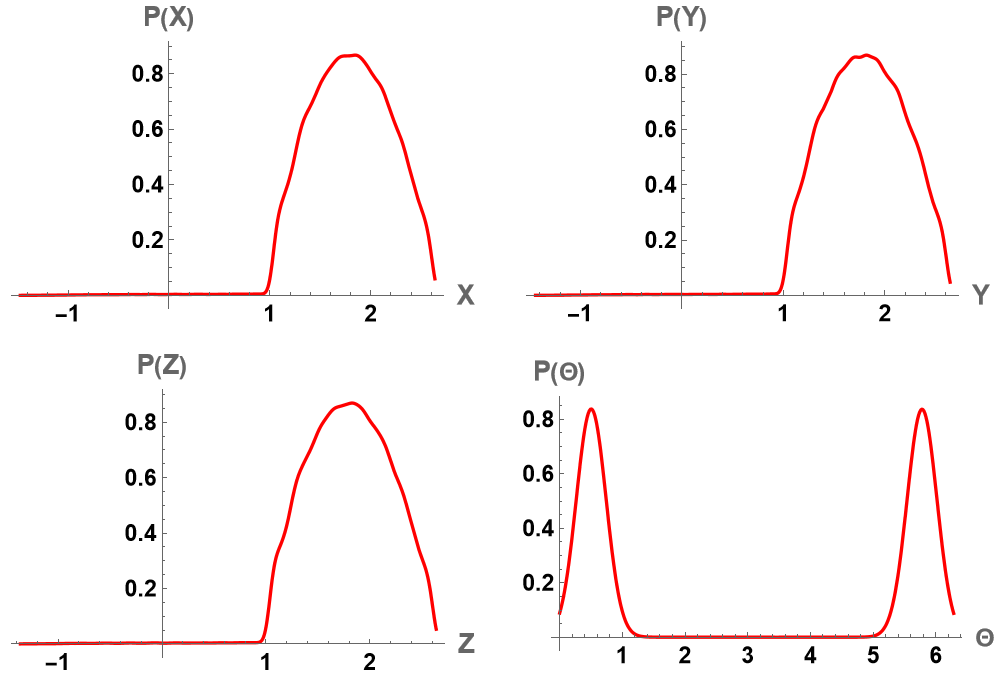}}
\hfil
     \subfigure[WCAPW]{\includegraphics[width=0.48\linewidth]{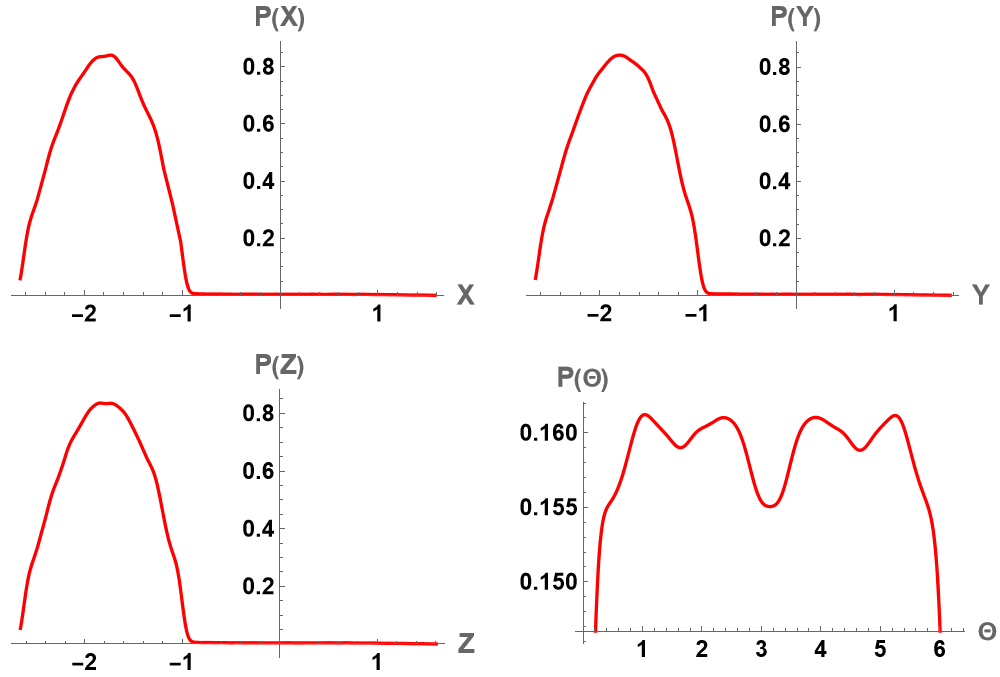}}  
\caption{  The probability distributions of space and phase variables  (a) DS (b) WCAPW  case.} 
 \label{F32}
\end{figure*}
\\
The distributions shifted to the right for the DS case and the left for the WCAPW case centred around $\pm 1.73$ the noted fixed points and maintained a bell shape. The angular contributions remain the same for DS, suggesting the preferred bi-stability. But there is a spread in phase distribution as before for WCAPW. The translatory shift in the position state space is due to the influence of the interaction forces and changed environmental conditions
[FIG.\ref{F32}].\\
\\
As the $b$ value decreased to $0.5$, the position entropy shows a steady state trend with a controlled movement pattern in the DS case. It retains certain randomness, as demonstrated by the relatively high value compared to the others. The phase value declines to zero very early, showing the impact of environmental conditions and there is an order disorder phase transition at a very low value of time$(t\sim 50)$. The phase angles become more predictable and underline the order in the phase of the system. The joint entropy plot shows that the overall randomness of the system does not increase over time, and there is a strong correlation between position state variables and phase. The mutual information plots suggest that relatively less independent variation in the state variables means one can predict to some extent given the others. So there exists a significant, crucial connection between the position and phase variables, underscoring the importance of our research. On the other hand, for $J=0.1,K=-1$ and $b=0.5$, the phase values are uniformly distributed and behave entirely randomly, adding an element of unpredictability to the system. The positional entropy is stable, but finite randomness persists. The mutual information curve suggests a strong coupling between spatial state variables and phase. The joint entropy underlines that the system is highly dependent rather than composed of independent variables. The entropy measures are shown in FIG.\ref{F33}

\subsection{Swarming states for conservative regime of Thomas oscillator [$b=0$]}
\begin{figure*}[hbt!]
    \centering     
    \subfigure[t=1200, TS]{\includegraphics[width=0.35\linewidth]{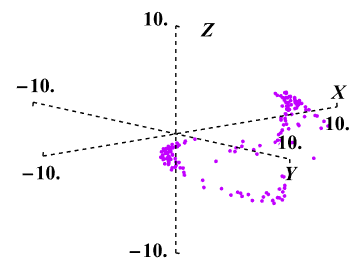}} 
    \hfil
    \subfigure[t=1220, TAS]{\includegraphics[width=0.35\linewidth]{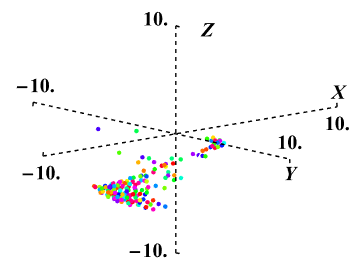}} 
\caption{  Swarming in 3D Space for the conservative regime of Thomas oscillator for randomly selected time values.  Simulations are done for $N=200$ for $T=1500 $ time units with $dT=0.01$ and $b=0$. We call the asymptotic swarming state as Turbulent Sync for $J=0.1$ and $K=1$ and turbulent Async for $J=0.1$ and $K=-ve$. } 
 \label{F5}
\end{figure*}

\begin{figure*}[hbt!]
    \centering
    \subfigure[TS]{\includegraphics[width=0.48\linewidth]{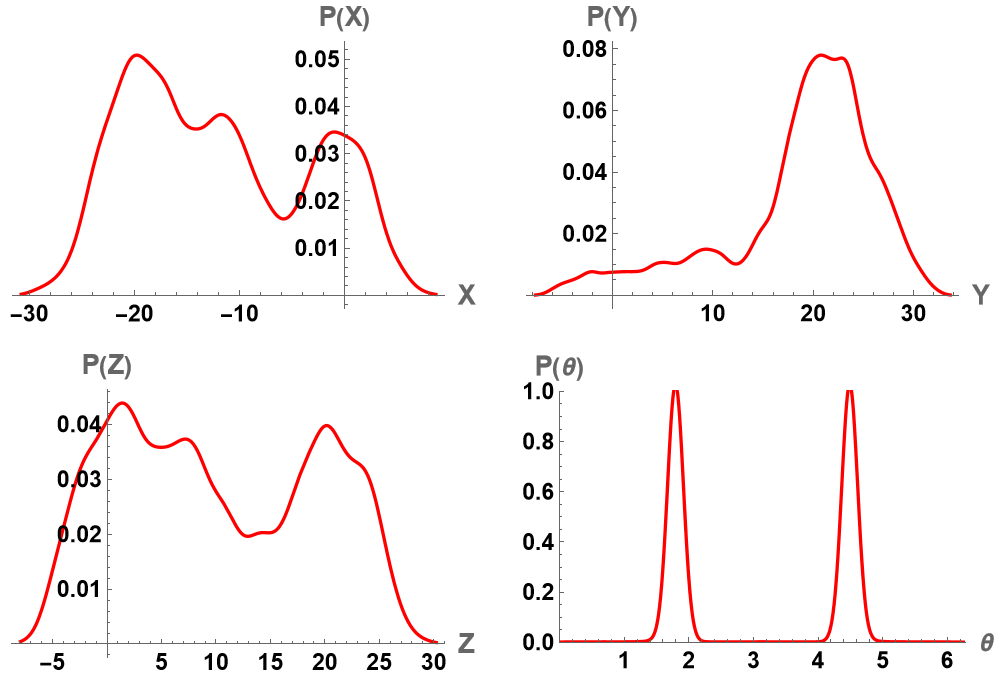}}
\hfil
     \subfigure[TAS]{\includegraphics[width=0.48\linewidth]{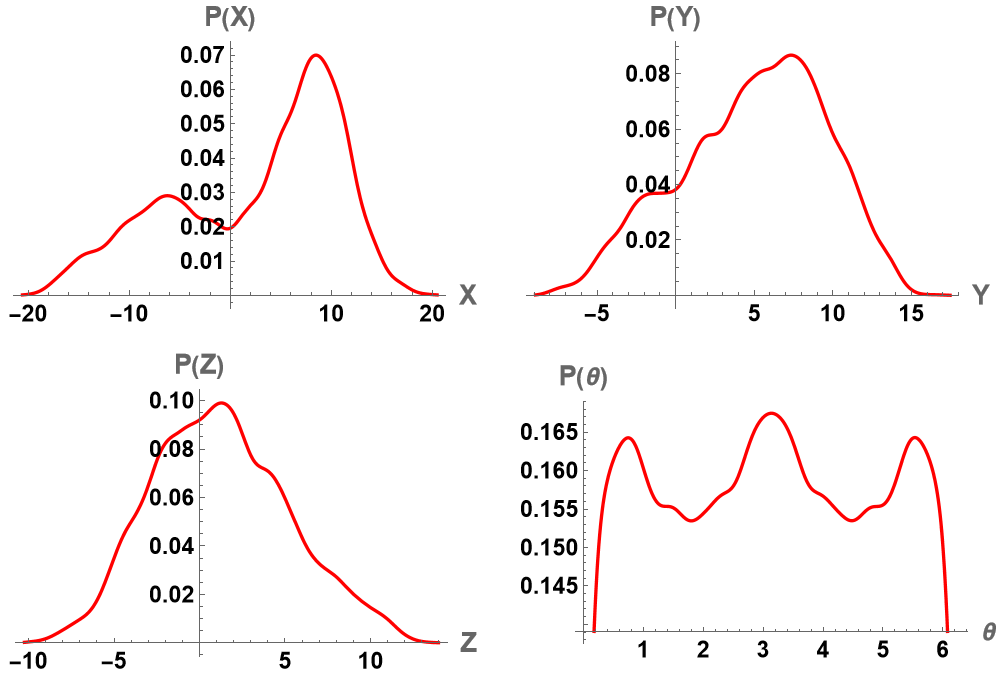}}  
\caption{The probability distributions of space and phase variables  (a) TS (b) TAS case. } 
 \label{F6}
\end{figure*}

\begin{figure*}[hbt!]
    \centering
    \subfigure[DS]{\includegraphics[width=0.48\linewidth]{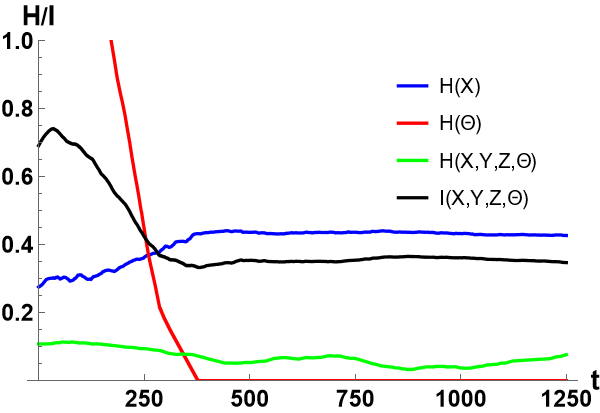}}
\hfil
     \subfigure[DAS]{\includegraphics[width=0.48\linewidth]{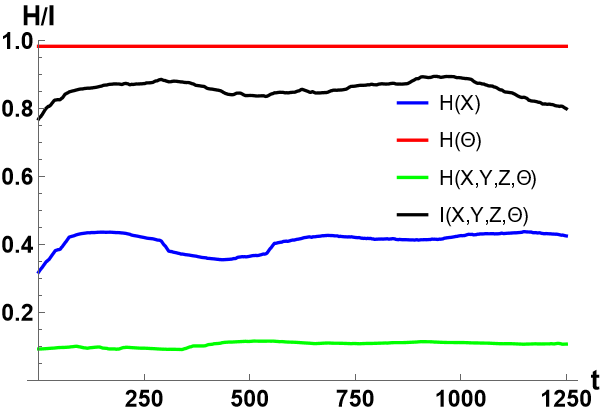}}  
\caption{Shannon entropy for space(Blue) and phase(Red) variables, joint entropy(Green), Mutual information(Black) plots for (a) TS (b) TAS } 
 \label{F7}
\end{figure*}

Without frictional damping ($b=0$), there is no mechanism to dissipate motion, leading to a dominance of inertial forces. This dominance results in collective behaviour characterized by large-scale swirling movements and intricate streaming patterns, known as active turbulence. For $b=0$, the system exhibits an infinite number of unstable fixed points, equally spaced along three mutually perpendicular directions at $n\pi$ where $n=0,\pm 1,\pm 2,.....$ forming a 3D grid. Due to the instability of these fixed points, particles explore phase space in circular, Zindler-type patterns, frequently migrating from one cluster to another around these fixed points. In this limit, $b=0$ reveals that the model effectively captures the physics of active turbulence. In this case, the Thomas system is conservative yet performs chaotic dynamics. The swarmalators exhibit intriguing behaviour as they move indefinitely in 3D space. Their motions involve frequent splitting and recombination, allowing them to explore a larger spatial region than others, resulting in a significantly larger mean square displacement. When $J$ is high, the cohesive force between like-phased swarmalators is substantial, forming a disc-shaped cluster. To distinguish these vigorous swarming behaviours, we introduce two specific states: Turbulant Synchrony (TS), where all oscillators move with the same phase, and Turbulant Asynchrony (TAS), where particles move with different phases but still exhibit phase-space correlation. Both states are characterized by finite values of $\mathcal{V}$ and large values of $\mathcal{D}$. The parameter ranges for these states are $(b=0,J>1,K=1)$ for TS and $(b=0,J>0,K=0,-ve)$ for TAS. These scenarios are illustrated in Fig.\ref{F5}.\\
\\ 
\\
Both in TS and TAS cases, there is a spread in probability for state space variables. Not only that these values are very low as compared with the previous two choices of the system parameter. This is an indication that the particles are executing chaotic motion and changes its state at each instant of time. Thus the broad spread in position state space variables suggests the system's transition between different dynamical states. For the TS case[FIG.\ref{F6}(a)], two sharp peaks indicate bi-stability in the phase dynamics. But this time they little nearer in phase values than before. TAS state shows almost unifrom distribution underlying maximum uncertainty in phase vales as depicted in FIG.\ref{F6}(b).
\\
\\
In this regime, a dynamic phase transition is observed. The system, initially in a chaotic phase, progresses towards a highly structured and correlated state. This transition, which occurs around $t\sim 250$, is marked by the decline of phase entropy to zero, indicating the emergence of phase synchronization and order. The positional entropy, starting at a low value, gradually increases and stabilises slightly above 0.4. This even distribution of spatial entropy over time leads to a significant increase in randomness, a crucial aspect of the system's evolution. The Mutual information, starting at a high value, indicates strong dependencies among the state variables. However, as soon as the phase entropy drops, mutual information also drops, meaning that phase and space dependencies weaken. Above $t\sim 500$, the mutual information stabilizes, indicating that some dependencies are still weaker than before. The joint entropy remains low but fluctuates, showing some structure yet with correlation among variables. In the turbulent asyc case, the randomness in phase is maximized with preserved spatial structure. This is the hallmark of a turbulent system where phase remains random but overall spatial structure is preserved[ref FIG.\ref{F7}].
\section{Discussion}
The Observed patterns, in general, represent lower entropy states than random disorders, showcasing the practical implications of the concept of entropy in understanding complex systems. Moderate entropy, a characteristic of complex systems, means the system has a myriad of potential configurations. \\
\\
Our research process, guided by the simulation results, has led us to a clear and confident conclusion: sharp changes in mutual information are strong indicators of potential phase transitions, synchronizations, and self-organizations. When its value stabilizes, it suggests the emergence of order or pattern formation. When positional entropy is low, particles form a structured pattern, similar to particle localization with a hint of randomness due to low variance. However, as its value increases, the tendency for particle diffusion also increases, leading to more intricate patterns in state space. This is precisely what we observe in circular-zindler-turbulent patterns. The shift of phase entropy from a high to a low value suggests the birth of global order and the loss of randomness. Conversely, when its value remains high and the remaining entropies fluctuate, the phase behaves like an external noise. In all cases, the joint entropy remains low, indicating self-organization and the emergence of structure- a collective behaviour phase transition.\\
\\
The formation of circular patterns, reminiscent of a biofilm, is a process intricately tied to the damping parameter. In the case of DS, particles with the same phase(internal property) clump together by minimizing energy, much like bacteria of the same species stick together to form a auto-aggregation as a survival policy or for food. Similarly, in the DAS state, we have particles with different internal properties that stick together, resembling an co-aggregation of bacteria- a symbol of coexistence. These two situations underscore the profound influence of the damping parameter on the overall dynamics. For instance, when the damping parameter is decreased from a high value to a low value, there is transition from a structured pattern to the turbulent motion of these aggregations, behaving exactly like an active fluid- a clear indication of adaptation.\\
\\
Compared with fluid dynamics, one can correlate the swarm's movement in state space in the TS case as a homogeneous flow since all the particles have the same internal property. Biomaterials like micro-fluid channels with Newtonian fluidity could be a close analogy from a technological point of view. On the other hand, in the TAS case, where we have different internal phases, we have a heterogeneous flow. A close analogy could be a chip device that uses multiple phases. The dynamics of the swarming system in the $b=0$ situation provide valuable insights that are crucial for designing innovative materials like self-propelling fluids and active polymers. Exploring the potential for self-healing biomaterials and bio-inspired soft robotics is integral to advancing these technological aspects and this research give valuable insights into these aspects.\\
\\
As far as the technological aspects of the research are concerned, plenty of applications are possible. We introduce phase, an internal property of the particle that could be spin, magnetic moment, or any other relevant physical quantity. If spin-like interactions are introduced, that can give insight into the dynamics of functional materials that can contribute to the development of advanced quantum devices, self-healing structures, responsive materials-new materials that mimic biological efficiency, adaptability and robustness. Self organizing structures highlighted in this research are well known for energy efficiency. The bi-stability observed in a phase probability distribution is crucial for developing programmable materials that can switch, adapt and respond to external stimuli in a controlled manner. Bi-stability is important because the system does not randomly fluctuate between states but instead follows a probability distribution that controls which probability is likely under specific conditions like temperature and light. Such materials are the cornerstone for developing magnetic skyrmions, deployable and reconfigurable structures. They are known as following generation information carriers in this modern era of computing.\\
\\
It is worth drawing a comparison between self-organizing particle swarms and AI, a comparison made possible through entropy. Just as in this simulation, where we began with random initial conditions, AI systems also start in a high-entropy state, with random structures and untrained data. Through the interaction of local units, the swarming systems self-organize and strive to minimize their entropy, all without a central organizer. Similarly, AI systems self-organize through the interaction of simple learning rules without the need for explicit programming and develop unexpected, often surprising, abilities beyond their training. In both scenarios, there is a dissipation of entropy. These complex systems can locally reduce entropy, harness energy, and self-organize into striking patterns, in the former case through interaction and evolution and in the latter through training and optimization.\\
\\
\section{Conclusion}
The agent-based swarming system, modelled through cyclically symmetric Thomas oscillators, mirrors the intricate dance of both natural and artificial self-organizing systems. In exploring this framework, we uncovered novel swarming states—DS, DAS, TS, and TAS—expanding our understanding of emergent collective behaviours. Beyond offering more profound insight into collective motion and adaptation, this study unravels the delicate and enlightening interplay between order and disorder through entropy-driven transitions. Leveraging Shannon entropy, joint entropy, and mutual information illuminates the pathways of information transfer among particles, revealing the mesmerizing emergence of structured patterns from apparent chaos. Despite entropy’s relentless drive toward disorder, complex systems defy this tendency, self-organizing into lower entropy states without external intervention—just as living organisms sustain order by dissipating entropy into their surroundings. This research advances our understanding of self-assemblies in colloids, complex fluids, and polymers and underscores the critical role of bi-stability in phase ordering, with profound implications for technological applications.

\bibliographystyle{apsrev4-2}  
\bibliography{reference}  

\end{document}